# Highly-crystalline 2D superconductors

Yu Saito[1], Tsutomu Nojima[2] and Yoshihiro Iwasa[1,3]*

[1] *Quantum-Phase Electronics Center (QPEC) and Department of Applied Physics,*

*The University of Tokyo, Tokyo 113-8656, Japan*

[2]*Institute for Materials Research, Tohoku University, Sendai 980-8577, Japan*

[3]*RIKEN Center for Emergent Matter Science (CEMS), Wako 351-0198, Japan*

*Corresponding author: iwasa@ap.t.u-tokyo.ac.jp

**ABSTRACT**

Recent technological advances in controlling materials have developed methods to produce idealized two-dimensional (2D) electron systems such as heterogeneous interfaces, molecular-beam-epitaxy (MBE) grown atomic layers, exfoliated thin flakes and field-effect devices. These 2D electron systems are highly-crystalline with less disorder in common, some of which indeed show sheet resistance more than one order of magnitude lower even in atomic layers or single layers than that of conventional amorphous/granular thin films. Here, we present a review on the recent developments of highly-crystalline 2D superconductors and a series of unprecedented physical properties discovered in these systems. In particular, we highlight the quantum metallic state (or possible metallic ground state), the quantum Griffiths phase in out-of-plane magnetic fields, and the superconducting state maintained in anomalously large in-plane magnetic fields, which were observed in exfoliated 2D materials, MBE-grown atomic-layer thin films and electric-double-layer (ion-gated) interfaces. These phenomena are discussed on the basis of weakened disorder and/or broken spatial inversion symmetry. These novel aspects suggest that highly-crystalline 2D systems are promising platforms for exploring new quantum physics and superconductors.

# 1. Introduction

Two-dimensional (2D) superconductivity has been developed over the last 80 years, and has kept providing a variety of quantum phenomena. Figure 1 shows the history of superconducting films toward the atomic thickness after 1980. In the early days of researches, the major methods of fabricating 2D superconductors were thermal evaporation and sputtering of metallic films, in which the most of basic concepts of 2D superconductors were established[1–5]. After 1980, physics of quantum phase transitions (QPTs) were intensively discussed and established in thin film superconductors ranging from ~0.3 to 10 nm in thickness, particularly with disordered structures mostly in the amorphous and granular forms (open symbols in Fig. 1)[6–13]. It is also noted that in 1990's a pioneering work on the fabrication of monolayer cuprate superconductors has been reported[14,15]. These researches were the dawn of new stages coming in the 21st century, in which a variety of new fabrication techniques, including molecular beam epitaxy (MBE) accompanied with surface/interface reconstruction process, mechanical exfoliation and field effect devices, have been introduced to the 2D superconductors researches (closed symbols in Fig. 1)[16–31]. As a result, the crystallinity was dramatically improved even in the atomic level thickness. With these newly emerging 2D superconductors, researchers are able to access the superconductivity in the 2D limit with highly crystalline forms. This is in sharp contrast with the 2D superconductors fabricated in the last century. The 2D superconductors nowadays are regarded as a platform for studying new physics as well as for searching novel and high-temperature superconductors.

In this article, we give a review of the recent development of 2D superconductor researches. In the next section, we first overview the history of 2D superconductors, and summarize the basic concepts created in this field together with the issues evolving to date. In sections from 3 to 7, we summarize the recently emerging 2D superconductors that include interface superconductors, highly-crystalline atomic layers grown by MBE, mechanically exfoliated 2D crystals, and ion-gated superconductors. In sections from 8 to 10, we discuss three major topics on the physical aspects of new 2D superconductors, the quantum metallic ground state (Sec. 8), the quantum Griffiths phase (Sec. 9), and the enhancement of the Pauli limit (Sec. 10).

## 2. History and issues of 2D superconductors

In 1938, Shal'nikovs first reported superconductivity in Pb and Sn thin films[1], which was the ground breaking work of thin film superconductors. Following this pioneering work, Buckel, Hilsh[2] and many researchers investigated different kinds of thin films made of soft metals and alloys[3–5]. The most sophisticated growth along this line is a quenched condensation method, where the elements of metals were deposited on surface of a substrate at low temperatures in an ultrahigh vacuum chamber immersed in liquid He. Those thin films were in amorphous or granular forms with the random arrangement of elements or the random orientation of their crystallites, and were therefore useful to investigate geometrical properties neglecting crystal orientation effects. For example, the angular dependence of upper critical field in thin film superconductors shows a cusp-like peak structure at the magnetic field direction parallel to the film plane[32]. This feature was well explained by the phenomenological theory based on the Ginzburg-Landau (GL) model. Tinkham pointed out in 1963[33] that the cusp-like peak is one of the clear characteristics of geometrical effect in 2D superconductors. According to the above scheme, 2D superconductors are defined as superconductors with their thickness thinner than the in-plane GL coherence length.

In addition, 2D superconductors shows many other interesting properties such as localization[34], transition temperature oscillation by quantum size effects[16,35,36], excess conductivity originating from fluctuation[37–39], Berezinskii-Kosterlitz-Thouless (BKT) transition[40–42] and QPTs at zero temperature[43,44]. In particular, the BKT transition, which is observed in the jump of the power-law exponent at zero current limit in current-voltage characteristic and the disappearance of ohmic resistance obeying the Halperin-Nelson scaling law, is treated as the evidence of 2D superconducting transition through the binding of vortex-antivortex pairs, as shown in previous works (Figs. 2a and b)[45].

One of the most important issues in 2D superconductors is the QPT[43,44], which has been intensively discussed to date. The superconductor-insulator transition (SIT) in metallic thin films was believed to occur only at zero temperature limit as a function of external tuning parameters, such as disorder (or film thickness), out-of-plane magnetic fields and carrier density (or electric fields), which determine the ground state of the system. The SIT in disordered systems can be divided in two-groups: one is originating from the amplitude fluctuation of order parameter, and the other includes the phase fluctuation. Fisher and others[43,46–48] suggested a "dirty-boson" model, considering that the SIT is caused by the quantum fluctuation of the phase and long-range Coulomb repulsions (interacting Cooper-pairs). They predicted that it is characterized by a universal sheet resistance (called as quantum resistance) $R_Q = h/(2e)^2$ at the transition point with $h$ the Plank constant and $e$ the elementary charge, which is independent of materials or systems, and the scaling behavior of temperature dependence of sheet resistance $R_{sheet}(T)$ against $|x - x_c|/T^{1/z\nu}$, where $x$ are $x_c$ are the tuning parameter and its critical value, and $\nu$ and $z$ the static and dynamic critical exponents, respectively. In the case of magnetic field induced SIT, this model can be understood from self-duality of Cooer pairs and vortices: the superconducting phase is described by a condensate of Cooper pairs with the localized vortices (vortex glass), while the insulating phase is a condensate of the vortices with localized Cooper pairs (Bose glass). Experimentally, some metallic films such as amorphous-Bi (Fig. 2c) shows a critical $R_{sheet}$ almost equal to $R_Q$ for disorder induced SIT[10]. The scaling collapse of $R_{sheet}(T)$ was also observed in the magnetic field induced SIT in films such as amorphous MoGe as shown in Figs. 2d and e[49]. However, it has been later pointed out that some others show different behavior[12], for example, an intervening quantum metallic state (possible metallic ground state) with a smaller sheet resistance than $R_Q$ between superconducting and insulating state[9,13,49–51]. Although many scenarios were proposed for this intervening metallic state, as discussed in the section 8, consensus is not yet achieved. A comprehensive review papers about SIT in conventional disordered 2D superconductors are published by Goldman and colleagues[44,52,53].

Another issue is the effect of spin-orbit interaction (SOI) on the superconducting state in a parallel (in-plane) magnetic field. In conventional 2D films in the dirty limit, the large parallel upper critical field exceeding the usual Pauli limit has been ascribed to the strong spin-orbit scattering which is contained in the pair-breaking parameter[54,55]. In the highly crystalline system with less disorder and long mean free paths, however, the spin randomizing effect of electron scattering as in the dirty model is not adequate and instead the effect of SOI should be considered using a different scheme, where it reflects the characteristic spin-split band structure or spin polarizations, originating from the broken spatial inversion symmetry. As we discuss in section 10 later, this causes an enhancement of the Pauli limit itself.

Owing to the recent technological advances and introduction of new approaches of thin films fabrications, there have been emerging a variety of 2D superconductors including interfacial superconductors[21,22,56,57], MBE-grown ordered-metal atomic layers[19,20], exfoliated single layer[25–27], electric-field-induced 2D systems[28,29,58], which we focus on in this review paper. In addition, chemical vapor deposition-grown atomic layers[59,60], heavy-fermion based super-lattices[61] and intercalated graphene[62–64] are also new classes of crystalline superconductors in the 2D limit. These new 2D superconductors are extremely thin in the atomic layer level and highly-crystalline in common, enabling us to address the issue of QPT in the minimal-disorder limit as well as to observe the novel superconducting state originating from the intrinsic nature of the materials such as crystal structure and/or spin-orbit coupling, possibly leading to a step forward to 2D topological superconductors[65].

3. Rise of interfacial superconductors

As mentioned above, until recently, the studies for SIT in 2D superconductors have been mainly performed using the deposited amorphous or granular thin films. In 2007, Triscone and Mannhart group was the first who succeeded in realizing superconductivity in a 2D electron system at the $LaAlO_3/SrTiO_3$(001) polarized interface fabricated by pulsed laser deposition (Figs. 3a and b)[21]. This was indeed a major breakthrough in the history of 2D superconductivity, because $LaAlO_3/SrTiO_3$ heterostructure have been known to show high electron mobility as compared with conventional metallic films[66]. After this success, they realized the electrostatic control of superconductor-metal-insulator transition by using a back gating technique (Fig. 3c)[67]. From their beautiful experiments, the superconducting phase diagram (the dependence of superconducting transition temperature $T_c$ on gate voltage $V_G$ or sheet carrier density) was found to be a dome-shape with the critical exponent of 2/3 in the vicinity of quantum critical point (at the threshold of the superconducting dome), which means that the superconducting-metal transition in $LaAlO_3/SrTiO_3$ is described by the (2+1)D XY model in the clean system[68]. As another important respects, the pseudo gap state and the Cooper pairs above $T_c$ are also observed with the tunnel spectroscopy[69] and single-electron transistor[70], respectively. As a similar system, the Mott-insulator-based $LaTiO_3/SrTiO_3$ interface was found to show superconductivity by Biscaras and colleagues[71–73]. One of the important discoveries in the electrostatically gated $LaTiO_3/SrTiO_3$ interface is the existence of multiple quantum criticality under magnetic fields, where the superconducting 2D electron system is driven to a weakly localized metal by passing through the double critical behavior, which is described by a clean and dirty (2+1)D XY model at low (high) and high (low) magnetic fields (temperatures), respectively[73]. An overview about interface superconductors is given by Gariglio, Triscone and colleagues[56,57].

Another important work is the observation of superconductivity at the cuprate-based interface between $La_{1.55}Sr_{0.45}CuO_4$ and $La_2CuO_4$ performed by Bozovic group (Fig. 3d)[22,74], who have pioneered the MBE growth of high-$T_c$ cuprate thin films[75]. While the two components are non-superconducting metal (M) and insulator (I) in the single-phase layers, respectively, the bilayers of them exhibit superconductivity with the deposition-sequence dependent $T_c$: 15 K in the I-M sequence and 30 K in the M-I sequence as shown in Fig. 3d. More importantly, they found that if the excess oxygen is doped in $La_2CuO_4$ of the M-I bilayer, $T_c$ exceeds 50 K, which is higher than the maximum value of the single phase $La_{2-x}Sr_xCuO_4$ and $La_2CuO_{4+\delta}$. Several mechanisms including electronic reconstruction at the interface, oxygen vacancies and interstitials have been proposed as possible origins of the improved $T_c$. However, there seems to be no conclusion reached[76].

He and co-workers were the first who reported the interfacial superconductivity between a topological insulator and an iron chalcogenide ($Bi_2Te_3$/FeTe), the heterostructure of which was fabricated by MBE[77]. They performed various analyses of transport data and confirmed that the superconductivity has the 2D nature with a thickness of ~7 nm, but the origin of superconductivity remains to be understood.

4. **Metal atomic layers grown by MBE**

Since the beginning of 2000's, the quality of metallic thin film superconductors (especially Pb) grown by MBE have continued to be improved to the extent that the intrinsic property in metallic ordered thin films can be investigated[16–18,78–80]. Oscillatory behavior of $T_c$ as a function of the number of monolayers was observed with reducing the thickness[16], which was accounted for in terms of the Fabry-Pe´rot interference modes of electron de Broglie waves (quantum well states) in the films[81,82]. In 2009, Qin and colleagues successfully observed superconductivity in Pb bi-layer crystalline films, which corresponds to a single quantum channel[19], and finally Zhang and co-workers did succeed in realizing highly-crystalline superconductivity in single-atomic-layer of ordered Pb ($T_c$ = 1.5 K) and In ($T_c$ = 3.2 K) ultra-thin films (Figs. 4a-f), which was confirmed by observing the superconducting gaps and vortices using scanning tunnel microscopy (STM)[20]. These atomic layers are the thinnest superconductors in existing systems (Fig. 1). After this success, Uchihashi et al.[83] and Yamada at al.[84] also performed transport measurements and confirmed superconducting transition in In and Pb ordered-atomic-layers (Figs. 4g and h). Particularly, Uchihashi and colleagues found that the atomic steps serve as strongly coupled Josephson junctions[83], where Josephson vortices exist at zero magnetic field[85]. Very recently, Matetskiy and colleagues succeeded in fabricating a 2D Tl-Pb atomic layer compound consisting of 1 monolayer Tl and 1/3 layer Pb on Si(111) substrate and in observing superconducting transition in zero and finite magnetic fields[86].

One of the most intriguing phenomena unique to these kinds of single atomic layer superconductors is the large Rashba spin splitting due to the spin-orbit interaction (SOI) with broken spatial inversion symmetry along the out-of-plane direction. In Tl-Pb ultra-thin films, the giant spin splitting (~250 meV) was indeed observed by angle-resolved photoemission spectroscopy (ARPES) measurements[87], which may lead to observe a number of fascinating phenomena. Okamoto group found the very robust superconductivity against the in-plane magnetic field in a Pb single layer on a cleaved GaAs(110) surface[88]. A very large in-plane upper critical field is suggestive of the enhancement of the Pauli paramagnetic limit by a factor of more than 4. In their discussion, the spin-orbit scattering effect in the dirty limit cannot account for the obtained results consistently. On the other hand, the results are explained in terms of an inhomogeneous superconducting state such as Fulde–Ferrell–Larkin–Ovchinnikov (FFLO) state or a helical state predicted for 2D metals with a large Rashba spin splitting[89–91].

Wang group in Peking University succeeded in growing Ga crystalline bilayers by MBE, and discovered the 2D superconductivity by transport measurements[92]. The same group discussed the unique magneto-transport properties observed in this system as mentioned below. Also, Hussain, Shen and Crommie group successfully realized the fabrication of MBE-grown NbSe$_2$ one-atomic-layer, and revealed the interplay between superconductivity and CDW formation in the 2D limit by ARPES and STM/scanning tunnel spectroscopy (STS) measurements[93].

5. **High-temperature superconductivity in FeSe atomic layers**

Xue Group reported an amazing work on a FeSe superconductor in 2012[23]. While $T_c$ for bulk FeSe is 8 K[94], they observed high temperature superconductivity in a FeSe single layer grown on SrTiO$_3$ substrates (Figs. 5a-c) with a superconducting gap of 20 meV at 4.2 K (Fig. 5d), which corresponds to $T_c \sim 80$ K, in the *in-situ* STM measurements. The onset $T_c$ of 5-unit-cell-thick FeSe with the capping layers is around 40 K in the *ex-situ* resistive measurement (Fig. 5e), which is still much higher than that of bulk. In this work, the authors mentioned that the bottom first unit cell contribute to the superconductivity. Later, it was reported that $T_c$ exceeded 100 K in FeSe single layer on Nb-doped SrTiO$_3$ substrate by the *in-situ* transport measurements[95]. From ARPES measurements, it is indicated that electronic structure is made of a single band crossing the Fermi level: the Fermi surface consists of electron pockets centered at the zone edge[96], and that an annealing process[97] and suppression of spin density wave[98] are necessary for the superconductivity. As the origins for the enhancement of $T_c$, the strain effect and charge transfer have been intensively discussed, and the former was found not to be promising[97–99].

Tsukazaki and Nojima group's results by using an electrochemical etching process in an electric-double-layer transistor (EDLT) configuration[100] provided the strong evidence that the electron doping and band bending due to the electric field or the charge transfer from the substrate play a crucial role in the enhancement of superconductivity in FeSe. According to their systematic studies of the effects of substrate ($SrTiO_3$ and MgO), thickness and carrier density, the high $T_c$ phase emerges even when the film is much thicker than the single layer under a strong electric field. They also found that a few-layer-thick FeSe film on $SrTiO_3$ exhibits superconductivity at about 40 K without gating. By contrast, FeSe on MgO superconducts only under gating possibly because of the absence of charge transfer from the substrate, in agreement with the previous results. A similar conclusion was reached by Takahashi group, who observed high temperature superconductivity in Potassium-deposited FeSe multilayer thin films[101].

## 6. Emergence of exfoliated 2D superconductors

Inspired by the discovery of graphene[102,103] and the subsequently growing researches of 2D materials in various approaches[104–106], researchers found that mechanical exfoliation is another useful method to create 2D superconductors with high crystallinity because in this method single crystal thin flakes are simply cleaved from bulk single crystals and transferred onto the substrate without crystal growth in vacuum chambers. If the deposition or epitaxial-growth method is called a bottom-up technique, this mechanical exfoliation can be regarded as a top-down method to fabricate 2D superconductors. Jiang and co-workers first succeeded in fabricating a single-layer $Bi_2Sr_2CaCu_2O_{8+x}$ (Bi2212) superconductor by cleaving them down to half-unit-cell thickness, where graphene was transferred on the top as the protection layer (Figs. 6a and b)[25]. Also, recent developments of the transfer technique to use hexagonal boron nitride (*h*-BN) as a substrate or cap-layer[107] allows us to address important issues in condensed matter physics. One of the typical exfoliated systems is a 2D $NbSe_2$ crystalline superconductor. Although Frindt made a pioneering work and successfully observed the thickness dependence of $T_c$ in ultra-thin NbSe$_2$[108], the real thickness was not so reliable because they estimated the thickness from resistivity. After many year's efforts by researchers[24,109], Cao et al. and Xi et al. demonstrated the superconductivity in a single layer NbSe$_2$ covered by graphene or h-BN[26,27] (Figs. 6c and d). In those studies, NbSe$_2$ was found to be conductive down to a monolayer and the evident signs of superconductivity were observed even in 1−3 layer samples, which shows interesting properties as discussed in section 8 and 10

### 7. Electric-double-layer transistor and electric-field-induced superconductivity

The field-effect transistor (FET) has been one of the most convenient tools to investigate the pure effect of carrier doping on various 2D materials without introducing unintentional disorder. In fact, this transistor structure has been utilized for a long time to modulate $T_c$ of conventional superconductor films as well as cuprate superconductor films[110–121]. Among them, Ahn and colleagues succeeded in remarkable control of $T_c$ and inducing SIT in ultrathin GdBa$_2$Cu$_3$O$_{7-x}$ films by using ferroelectric oxide gate[118]. After this success, Goldman group realized the electrostatic modulation of SIT in amorphous Bismuth thin films[120], and Triscone and Mannhart group succeeded in continuous switching from superconducting state to insulating state in LaAlO$_3$/SrTiO$_3$[67]. However, the bottleneck of this conventional FET structure with solid gate dielectrics, which might drag down the research of electric-field-effect control of superconductivity in many materials, is the magnitude of controllable carrier density. The reachable carrier density by the field effect is limited because of the dielectric breakdown, and usually is not enough to induce superconductivity in originally insulating materials and to control $T_c$ in a wide range.

To solve this bottleneck, a different type of electrostatic doping technique using a liquid gate, which is realized in the EDLT configuration, has been developed[122–124]. The basic concept for carrier doping in EDLT is similar to that in conventional FET structure, where an ionic liquid or electrolyte is employed as a gate medium instead of solid gate dielectric (Figs. 7a and b; Figure 7a shows a typical optical image of MoS$_2$ thin flake device). According to this replacement of gate dielectric materials, the EDLT enables us to produce a large electric field over 10 MV/cm and to accumulate the extremely high carrier densities over $10^{14}$cm$^{-2}$ by forming an EDL with a ~ 1-nm in thickness at the surface, which can be regarded as a nanometer-gapped capacitor. While the function of the prototype transistor with electrolyte was examined in very early studies in some materials including organic semiconductor[125] and cuprate superconductors[126], Iwasa and Kawasaki group has improved this technique and first succeeded in realizing the electric-field-induced superconductivity on the surface of SrTiO$_3$[28], which is originally a band insulator. After this success, the EDLT has become a promising tool for the search of superconductivity, as exemplified by gate-induced or tuned superconductivity in KTaO$_3$[127], quasi-2D layered ZrNCl[29], transition metal dichalcogenides (TMDs)[30,31,128–130] and cuprate thin films[131–134] in addition to SrTiO$_3$[28,135,136]. In particular, ZrNCl and MoS$_2$ were investigated in detail as for $R$-$T$ curve at different gate voltages (Fig. 7c, d) and $T_c$ vs carrier density (Figs. 7e, f) were found to exhibit a dome-shaped phase diagram (Figs. 7e, f)[30,137]. These phenomena remind us of the similar dome-like phase diagram in high-$T_c$ superconductors such as cuprates and iron pnictides, as well as that of low $T_c$ in SrTiO$_3$. These dome-like phase diagrams suggest that there seems to be a commonality of all the superconductors induced by carrier doping, irrespective of the electronic states in the parent materials. Therefore, understanding the phase diagram of 2D systems might be one way to establish the basic concept of high-$T_c$ superconducting mechanism.

A great advantage of the EDLT is its easy accessibility to a wide range of materials because it is not necessary to grow the solid dielectric thin film on the top of channel materials in contrast to the usual FETs. Indeed, the methods of EDLT are now applied to the studies for not only superconductivity but also other physical phenomena, such as ferromagnetism, metal-insulator transitions, tuning of SOI and valleytronics[138–143].

## 8. Quantum metallic state in highly-crystalline 2D superconductors

Emergence of the highly-crystalline 2D superconductors as introduced in the previous sections allows us to investigate the intrinsic behavior of quantum phase (ground state) in a minimal-disorder limit, which is one of the long-standing and important issues of 2D superconductivity to be addressed in conjunction with the SIT. In this section, we discuss the recent two works on the observation of quantum metallic state in ion-gated ZrNCl[137] and NbSe$_2$ bilayer superconductors[144].

ZrNCl is a layered band insulator, which was found to be superconducting both in the bulk and at the surface by alkali-metal intercalation[145–148] and electrostatic doping using the EDLT[29], respectively, with the similar maximum $T_c$ of approximately 15 K as shown in Fig. 7e. In particular, owing to the weak van der Waals force between the layers, nanometer-thick ZrNCl crystals with an atomically flat surface, which is indeed suitable for the EDLT device, can be cleaved *via* mechanical exfoliation. Saito and colleagues reported a comprehensive study on magneto-transport properties on ionic-liquid gated superconducting ZrNCl[137]. Figure 8a shows the typical $R_{sheet}(T)$ curves for $V_G$ = 6.5 V ($T_c$ = 15 K) under different out-of-plane magnetic fields. In this geometry, $T_c$ is substantially suppressed by the magnetic field up to 2 T, whereas it does not change so much even at 9 T in the in-plane magnetic field geometry. In addition, as shown in Figure 8b, the angular dependence of the upper critical field, $H_{c2}$ ($\theta$) reveals a cusp-like peak at $\theta$ = 90° (Fig. 8b, inset). Here, $\theta$ represents the angle between the *c*-axis of ZrNCl and applied magnetic field directions, and $H_{c2}$ is defined as the magnetic field where $R_{sheet}(H)$ becomes 50 % of the normal state resistance, $R_N$, at 30 K. Figures 8a and b clearly prove that the superconductivity in the ZrNCl-EDLT is in the extremely 2D limit.

Indeed, they succeeded in fitting the data of $H_{c2}(\theta)$ in Fig. 8b by the 2D Tinkham model but failed by the three-dimensional effective mass (GL) model. Assuming that the $H_{c2}(T)$ both for the in-plane and out-of-plane conditions are determined by the orbital limit as described in the GL model, the effective superconducting thickness is estimated to be $\cong 1.8$ nm, which is smaller than one-unite-cell thickness of 2.8 nm. This assumption might be questionable for the in-plane magnetic field because the Pauli limit might play a significant role, meaning that $H_{c2}(90°)$ in the orbital limit positions at a higher value. Therefore, the value of 1.8 nm should be regarded as the upper limit of the real superconducting thickness.

The most remarkable feature to be noticed in Fig. 8a is that $R_{sheet}(T)$ is dramatically broadened even at a small magnetic field of $\mu_0 H = 0.05$ T. This indicates that the vortices are easy to move owing to the extremely weak pinning effect at the atomically flat surface, resulting in the ohmic dissipation. It is noted that this broadening behavior consists of two kinds of dissipation process, appearing at high and low temperatures. As shown in the Arrhenius plot of Fig. 8c, $R_{sheet}(T)$ exhibits an activated behavior at high temperatures just below $T_c$ described by a function form $R_{sheet}(T,H) = R'\exp(-U(H)/k_B T)$, where $k_B$ is Boltzmann's constant, as indicated by dashed lines. The logarithmic magnetic field dependence of the extracted activation energy $U(H)$ (Fig. 8d), and the relation between $U(H)/k_B T_c$ and $\ln R'$ (Fig. 8d, inset), are in good agreement with the thermally-activated collective vortex-creep model in a 2D superconductor with $U(T,H) \propto (1-T/T_c)\ln(H_0/H)$. This indicates that the dissipation process occurs due to the plastic motion of dislocation pairs in 2D vortex lattice[149]

At low temperatures, by contrast, each $R_{sheet}$–$T$ curve deviates from the thermally-activated behaviour, and then shows saturation at a finite resistance down to the lowest measurement temperature (= 2 K), being suggestive of a quantum metallic state (or possible metallic ground state) in this system. There was no sign of transition from the metallic phase to the superconducting vortex solid (vortex glass or Bragg glass with $R_{sheet} \to 0$) as seen in Bi$_2$Sr$_2$CaCu$_2$O$_8$ single crystals[150] and YBa$_2$Cu$_3$O$_y$ ultra-thin films[15]. A similar phenomenon has been observed in weak pinning systems such as amorphous MoGe[50,151,152], MoSi[153], Ta[13,154–156] and Bi[51] ultrathin films. To explain this quantum metallic state, three paradigms were proposed. Galitski and colleagues propose a model of magnetic-field induced-motion of vortices interacting with so-called spinons (vortex metal)[157], which well accounts for the large peak in $R_{sheet}(H)$ above the critical field of SIT observed in amorphous InO$_x$ films[158]. Shimshoni et al. considered the vortices motion though the quantum tunneling, where the dissipation occurs due to gapless Fermi liquid (fermionic) excitations in the vortex cores[159]. This is indeed in good qualitative agreement with the experimental results on the amorphous MoGe thin films obtained by Mason and Kapitulnik[151]. Third possibility is a "Bose metal", a Cooper pairs forming gapless non superfluid liquid due to the zero-point motion of vortices at zero temperature, which was proposed by Das and Doniach[160,161]. After this proposal, a similar scenario, the coupling of bosonic degrees of freedom to the excitation of the glassy phase (phase glass), was suggested by Philips and others[162–164].

In the ZrNCl-EDLT, Saito et al. suggested that the most plausible origin of the metallic state at low temperatures is quantum tunneling of vortices (quantum creep) in the model of percolating weak-link-network consisting of superconducting and normal (vortex liquid) puddles coupled to a dissipative fermionic bath,[159] which is likely in the system with small $R_N$ ( ~1/50 of $R_Q$) and weak but finite disorder:

$$R_{sheet} \sim \frac{\hbar}{4e^2} \frac{\kappa}{1-\kappa}, \qquad \kappa \sim \exp\left\{ C \frac{\hbar}{e^2} \frac{1}{R_N} \left( \frac{H - H_{c2}}{H} \right) \right\} \qquad (1)$$

where $C$ is a dimensionless constant. As seen in Fig. 8e, the $R_{sheet} - H$ relation at 2 K is well fitted by eq. (1) up to 1.3 T, which indicates that the model of dissipation process by quantum creep holds for a wide range, although $C$ is much smaller than the order of unity. This might be because the ZrNCl shows very low normal state resistance ($k_F l \sim 100 \gg 1$, where $k_F$ and $l$ are Fermi wavelength and mean free path, respectively) compared to with conventional metallic thin films, and also the quantum creep model assumes only dirty systems.

Above 1.3 T, $R_{sheet}$ exhibits the $H$-linear dependence, which corresponds to the crossover from the vortex creep to the vortex flow motion at 1.3 T. At this magnetic field, $U(H)$ for the thermally activated creep approaches zero (Fig. 8d), implying that the pinning or the elastic potential effectively disappears at high magnetic fields. Based on the above observations, Saito et al. obtained the vortex phase diagram ($H$-$T$ phase diagram) of ion-gated ZrNCl shown in Fig. 8f. Most part of vortex phase diagram is occupied by a metallic phase, suggesting the quantum metallic phase due to the collective quantum creep of vortices is the ground state under magnetic fields. Importantly, this metallic state appears when the magnetic field of 0.05 T (only 1/40 of $H_{c2}$) is applied and therefore the zero-resistance region is indiscernible. The metallic ground state has been intensively discussed using amorphous or granular films but was observed only at magnetic field high enough to destroy the stable zero-resistance state. Such a fragile zero-resistance state demonstrated in Fig. 8f is attributed to the extremely weak pinning due to the highly crystalline nature coupled with two dimensionality.

A similar quantum metallic state in a wide range has also been reported by in a 2D crystalline NbSe$_2$ bilayer by Tsen et al.[144] The normal state $R_{sheet}$ of bilayer is 75 Ω, which is comparable to the lowest value of the gate-induced ZrNCl superconductor[137]. They investigated the magnetotransport properties and found the temperature-independent resistance saturation at low temperatures. They discussed this phenomenon in terms of a "Bose metal" model based on the analysis in which they found that the power law field dependence of resistance as expressed by the following equation[162]:

$$R_{\text{sheet}} \sim (H - H_{c0})^{2\nu}. \tag{2}$$

Here, $H_{c0}$ is the characteristic field of true zero resistance state. As shown in Fig. 9a, the magnetoresistance data are quite well fitted by Eq. (2), where $2\nu$ changes from 1 at high temperature (thermally activated vortex flow) to 3. From these observations, they considered that the metallic state at low temperatures can be attributed to Bose-metal state as shown in Fig. 9b. Although the current interpretations of quantum metallic states are different between the gate-induced $ZrNCl$ and the bilayer $NbSe_2$, both the models are based on the quantum motion of vortices. In the absence of $H_{c0}$ in Eq. (2) or in the magnetic fields far from it, the collective quantum creep of vortices might be the case, while, on the other hand, in the presence of pinned vortex state, the Bose metals might describe the dissipation near $H_{c0}$. Both the processes seem to appear depending on the tiny difference of pinning strength, which is very weak but not zero in the systems discussed in this section, and they are connected with each other continuously with increasing pinning strength or magnetic field. In any case, the metallic ground state may be the intrinsic nature of highly-crystalline 2D superconductors with extremely weak pinning.

The wide range of quantum metallic state under magnetic field observed in EDLT and exfoliated 2D superconductors has not been clearly observed in oxide interfaces[21,165] or MBE-grown ordered-metal atomic layers[86]. This may be because of the unintentional deficiencies, grain boundaries or surface atomic steps, which are inevitably included during the film growth process and can become strong pinning centers even in highly-crystalline materials. In the strong pinning limit, the possible ground state is a vortex glass phase with $R = 0$, leading to the suppression of the intervening metallic state. Recently, in In thin films, the atomic steps are found to behave as the strongly coupled Josephson junctions[83]. These can be strong pinning centers when the applied current is along the steps.

At the end of this section, we address the quantum metallic state without magnetic field observed in gated $LaTiO_3/SrTiO_3$ interface[72]. In this particular case, vortex dissipation model cannot explain this phenomena, but, according to the authors, the spatial inhomogeneity in the interface instead causes percolating network or phase separation of the superconducting state and the normal state, leading to the finite resistance state even at zero magnetic field.

Recent observation of the critical exponent of 3/2 in this oxide interface system also supports this scenario[68].

## 9. Quantum Griffiths singularity of superconductor-metal transition

In the previous section, we discussed the metallic ground state induced by small magnetic fields in highly-crystalline 2D superconductors at very low temperatures. As the magnetic field further increases, this state involving incoherent Cooper pairs is eventually broken and then weakly localized normal metal without Cooper pairs or insulating state appears through QPT. In treating this QPT, we should carefully consider how the effect of disorder evolves with the length scale (coarse graining) in the systems with quenched disorder such as impurities, dislocations and grain boundaries. If the average disorder strength decrease or approach a constant under coarse graining, corresponding to the Harris criterion $d\nu \geq 2$ with the space dimension $d$ and the static exponent $\nu$[166], the QPT can be treated experimentally in the similar scheme to a conventional clean quantum critical point (QCP) with a constant $z$ characterizing the time scale of the QPT. On the other hand, if the average disorder strength increases and diverges under coarse graining in the system with $d\nu < 2$ and weak random disorder, the clean QCP is considerably modified by the disorder. In this case, the probability distribution of the order parameter becomes very broad in an infinite size system, and as a result, there emerges an exponentially small but nonzero probability for the existence of large ordered regions (Fig. 10a). Such rare regions can be locally ordered in one phase while the whole system is in the other[167–169]. Griffiths[167] showed that these locally ordered islands (rare regions) play an essential role in determining the QCP and the dynamics of its critical region. This dynamics results in the so-called Griffiths singularity characterized by an infinite-randomness QCP and the divergence of $z$ toward it.

The quantum Griffiths singularity has been the most dramatically illustrated in the quantum random transverse-field Ising model (Fig. 10a)[170,171], and widely studied by theorists but with limited experimental evidence in three dimensional ferromagnetic metals[172,173]. It has been a challenging issue to observe it experimentally in 2D superconducting systems, although there are two related studies which report unusual critical behavior such as multiple quantum critical behavior of a superconductor-metal transition (SMT)[73] or a two-stage magnetic field-tuned SIT[174]. Here, we introduce recent experimental works, which for the first time showed the clear feature of the quantum Griffiths singularity[175,176].

Wang and Xie group investigated the magneto-transport properties in Ga crystalline trilayer thin films[175] grown by MBE technique, showing the BKT transition in zero magnetic field (Fig. 10b), down to ultralow temperature regimes, and found that multiple crossing points in resistance versus magnetic field curves as shown in Fig.10c. This behavior is in marked contrast to the conventional single QCP as is frequently observed in amorphous metallic thin films, for instance, shown in the inset of Fig. 2d. They performed finite size scaling analyses at different temperature intervals and found that critical exponent $z\nu$ depends on temperature showing diverging behavior toward QCP (Fig. 10d).

This anomalous scaling behavior is consistent with the quantum random transverse field Ising model, which shows the activated scaling with continuously varying dynamical exponent when approaching the infinite-randomness QCP. Considering $\nu$ of 0.5 theoretically predicted for the SMT in a clean 2D superconducting system[176], the Harris criterion is violated with $d\nu = 1$. In this case, the introduction of finite quenched disorder can lead to the infinite-randomness QCP. Following this scenario, the product of the critical exponents $z\nu$ is expressed by the activated scaling law[177–180]:

$$z\nu \approx C(H - H_c^*)^{-\nu\Psi} \qquad (3)$$

with constant $C$ and the 2D infinite-randomness critical exponents $\upsilon \approx 1.2$ and $\Psi \approx 0.5$[181,182]. This is in good agreement with their experimental data, indicating the existence of the quantum Griffiths singularity as an infinite-randomness QCP. Similar phenomenon was observed in LaAlO$_3$/SrTiO$_3$ (110) interface system by the detailed magnetotransport studies in the ultra-low temperatures[183].

In type-II superconductors, especially in the disorder-free case, the vortex lattice state melts by thermal fluctuation at a melting temperature $T_M$, and changes to the vortex liquid phase[38,184]. However, once including weak quenched disorder, the effect of which is not negligible in the system, the long-range order of a vortex lattice can become also unstable against the fluctuations caused by disorder even at much lower temperature below $T_M$. As a result, with increasing magnetic field the vortex lattice phase evolves into a vortex glass-like phase composed of spatially separated superconducting islands (Griffiths regions or rare regions) connected *via* long-range Josephson junction arrays (or weak-linked superconducting and normal puddles). Because in the limit of low temperature the fluctuation effect of the order parameter by quenched disorder become much more predominated than the thermal fluctuation effect, the superconducting islands can behave as the Griffiths region or rare region.

They attributed the absence of quantum Griffiths singularity in the previous studies to the thermal fluctuation effect, which can smear out the inhomogeneity caused by quenched disorder. In addition, we speculate that another possible reason of the absence is the bad crystallinity of previous 2D superconductors. In the early works, the main target to investigate critical exponent is amorphous/granular thin films, which include large amount of quenched disorder, in other words, the concentration of disorder was high and homogeneous. Indeed, there have not been clear multiple critical points in most of the studies on amorphous granular, or polycrystalline thin films. For example, although Yoon group measured amorphous Ta films with a $T_c$ of 0.23 K down to 0.053 K[13], they did not observe the quantum Griffith phase, implying the absence of rare large ordered region in amorphous thin films. In such homogeneously disordered systems, the effect of quenched disorder can be averaged out under coarse graining without violating the Harris criterion, resulting in the conventional QCP above which the localization of preformed Cooper pairs simply occurs. On the other hand, in crystalline systems, ordered regions and weakly disordered regions can coexist. This situation is close to the 2D lattice of spins as shown in Fig.10a, which satisfy the condition of the emergence of the quantum Griffiths singularity at low temperatures. Therefore, the quantum Griffiths phase may be a sign of less disordered and highly-crystalline 2D superconductors.

As mentioned in the previous section, the wide-range spatial inhomogeneity of the order parameter can be also observable in the system with the electronic phase separation as reported in oxide interfaces[68,73,185], where the superconducting region with the higher carrier density is surrounded by metallic matrix with the lower density. This situation results in a different criticality from that of the usual SIT, but the key parameter, which distinguish the phase separation behavior from the quantum Griffiths phase, may be the critical exponent $z\nu$ depending on magnetic field (temperature). In the former case the value approaches a constant value, but in the latter case it shows a diverging behavior toward $H_c^*$ according to Eq. (3). For a further understanding of such Griffiths phenomena, it is necessary to examine the QPTs in other highly-crystalline 2D systems and discuss them including previous data of conventional thin films such as $InO_x$.[11,158,186,187]

## 10. Enhanced upper critical field in noncentrosymmetric 2D superconductors

Finally, we address the robust superconducting state against magnetic fields due to the anomalous effect of SOI in recently emerging highly-crystalline 2D superconductors without spatial inversion symmetry. In 2D superconductors, the upper critical field determined by the orbital limit becomes infinitely large, and instead the Pauli limit can be dominant in the in-plane magnetic field geometry. In early stage works, the parallel upper critical field $H_{c2}^{\parallel}$ larger than the usual Pauli limit $H_{P}^{BCS} \sim 1.86T_c$ based on Bardeen-Cooper-Schrieffer (BCS) theory for weak-coupling superconductors had frequently been attributed to the effect of strong spin-orbit scattering causing the spin randomization in high-Z elements (a-Bi or a-Pb) or layered materials with large atomic SOI[54,55], because these metallic thin films superconductors were dirty systems with very short mean free paths compared to coherence lengths in most cases. However, once the system has less disorder and then enter the clean regime, the spin-orbit scattering effect becomes weak, and instead the effect of spin-momentum locking based on SOI can much contribute to enhancement of the $H_{c2}^{\parallel}$, especially in noncentrosymmetric superconductors. So far, the experimental works related to large upper critical field enhanced by spin-momentum locking have been limited to the system with Rashba-type SOI as is represented mainly by heavy fermions[91,188–190].

Superconducting atomic layers based on TMDs are highly-crystalline 2D systems with Zeeman-type SOI (also called as Ising SOI)[140,191,192]. In the monolayer limit, the in-plane inversion symmetry in the crystal is spatially broken as shown in Fig. 11a. which causes a special type of SOI producing out-of-plane spin polarization together with effective Zeeman fields, namely the Zeeman-type spin polarization, leading to the spin split band at the K valleys[191,193] (Fig. 11b) at zero magnetic field. This is in marked contrast to the Rashba-type spin splitting with the helical spin texture, and is the intrinsic nature of TMDs monolayer that originates from its $D_{3h}$ crystal symmetry (Fig. 11a). The spin-valley locking was experimentally confirmed by for the valence band of $MoS_2$ by spin-resolved ARPES[194]. Recently, the enhanced $H_{c2}^{\parallel}$ through this mechanism was observed in two systems; ion-gated $MoS_2$[195,196] and $NbSe_2$ bilayer[197].

Saito and co-workers investigated temperature dependence of $H_{c2}^{\parallel}$ of MoS$_2$ in an EDLT configuration by using pulsed high magnetic fields up to 55 T down to low temperature, and observed enhanced $H_{c2}^{\parallel}$ far above the $H_{P}^{BCS}$ [195]. It is noted that $H_{c2}^{\parallel}(T)$ increases with decreasing temperature and eventually saturates approximately 52 T at 1.5 K, which is more than four times larger than $H_{P}^{BCS}$ = 12 T, as is shown in Fig.11d. Similar experimental results were obtained by Ye group[196].

Both groups first discussed this anomalous enhancement of $H_{c2}^{\parallel}$ in terms of the spin-orbit scattering effect by using the microscopic Klemm–Luther–Beasley (KLB) theory[198], which is traditionally applicable to dirty-limit layered superconductors with strong SOI. In fact, the experimental data themselves are well fitted by the KLB theoretical curve. However, the value of spin orbit scattering time estimated from the fitting curve turns out to be much shorter than that of total scattering time obtained based on the transport data. This unphysical result implies that application of the KLB theory is not appropriate, so that the spin-orbit scattering effect is not a dominant origin for the enhanced $H_{c2}$ in ion-gated MoS$_2$[195],[196].

An alternative is the intrinsic mechanism considering the highly-crystalline nature of MoS$_2$ conduction plane, that is, the spin-valley locking as shown in Fig. 11b. According to the first-principles-based band calculations, the conduction bands at the K and -K valleys are spin-split with almost fully spin polarization in out-of-plane directions. In the absence of magnetic field, the spin polarizations between K and –K valleys are opposite due to the time reversal symmetry. The Fermi energy $E_F$ and the spin splitting at $E_F$ were obtained as 150 meV and 13 meV, respectively, indicating that the electrons feel the internal out-of-plane magnetic field about 100 T[195]. Fig. 11c displays a schematic of the Fermi surface with the spin polarization where electrons in the K and –K valleys are spin-polarized predominantly upward and downward, respectively. To form a singlet pairing state with zero center of mass momentum, Cooper pairs should be formed between K and –K valleys. Such inter-valley pairs is protected by the internal out-of-plane magnetic field, meaning that, unless the external in-plane magnetic field exceeds the internal magnetic field, Cooper pairs remain robust. This is the qualitative mechanism of the anomalous enhancement of $H_{c2}^{\parallel}$ or Pauli limit owing to spin-valley locking, which is also referred to as Ising superconductivity or Ising pairing[195–197].

It is worth noting that the in-plane Rashba-type component originating from the application of an out-of-plane electric field may exist. In this case, due to the presence of both the Zeeman type SOI and small Rashba-type SOI, a FFLO state (or a helical state)[89–91] with $s+f$-wave symmetry, can be realized. The relation between crystal symmetry, $g$-vectors and pairing-symmetry is discussed in the Frigeri's Ph.D. thesis[199].

For more quantitative understanding, Saito and coworkers numerically estimated the realistic Pauli limit field of the present system based on the tight binding model, assuming the $s$-wave singlet Cooper pairs and taking into account both the Zeeman- and the small Rashba-type SOI[195]. For the values of SOI, the former is obtained in the first-principles-based band calculations and the latter is used as a fitting parameter. Considering only the Zeeman-type SOI, the Pauli limit is considerably enhanced as it is larger than 70 T at $T = 1$ K. Once the small Rashba-type SOI is introduced, Pauli limit is substantially suppressed, because the in-plane polarized spin components originating from the small Rashba SOI is much more susceptible to an external in-plane magnetic field than the intrinsic Zeeman-type SOI. The best agreement is obtained when we assume 10 % Rashba component of the total spin polarization as shown in the sold line in Fig. 11d. Because the first principle calculation predicts only 2 % of Rashba component, the agreement is still in a semi-qualitative level. Nevertheless, the above result strongly indicates that spin-valley locking is the major origin for the enhancement of $H_{c2}^{\parallel}$.

Xi and co-workers also observed large in-plane $H_{c2}$ in NbSe$_2$ atomic layers[197]. Bulk $2H$-NbSe$_2$ is a conventional type-II superconductor with a $T_c$ of 7 K. Monolayer NbSe$_2$ can be viewed as to be equivalent to heavily hole-doped monolayer MoSe$_2$[200,201]. The Fermi surface is composed of one pocket at the Γ point and two pockets at the K and the -K point of the Brillouin zone[201], which is the different point from the semiconducting TMDs such as $MoS_2$. In the monolayer limit, a large spin splitting appears at the K valleys. They observed the robust superconducting state of $NbSe_2$ mono-, bi-and tri-layers in in-plane magnetic fields applied up to 31 T as shown in Fig. 11e. In particular, the $H_{c2}^{\parallel}$ exceeds the Pauli limit by a factor of 4 even near $T_c$, suggestive of the huge enhancement of the Pauli limit owing to the spin-valley locking. Furthermore, they attributed the observation of the large $H_{c2}$ observed in bi- and tri-layers $NbSe_2$ to the spin-layer locking[202], where the electron spins are locked to each individual layer instead of being locked to the momentum, which also protects superconductivity under a parallel magnetic field.

These results thus indicate that highly-crystalline and clean 2D superconductors provide a new platform for investigating unprecedented exotic properties owing to broken inversion symmetry combined with strong SOI, which may lead to discoveries of 2D topological superconductors. Indeed, nodal topological superconductivity, which has nodal points connected by Majorana flat bands, is predicted in monolayer $NbSe_2$.[203]

## 11. Summary

We have reviewed recently emerging 2D superconductors. The common feature of these 2D superconductors is their high-crystallinity that makes them distinct from the conventional 2D superconductors whose structures are granular or amorphous. Such highly-crystalline 2D superconductors will be a new materials platform for superconductivity not only to understand low dimensional specific properties but also to realize high $T_c$ superconductors.

So far we have not argued the values of $T_c$, despite their importance. It has been believed that $T_c$ in the thin film superconductors decreases with decreasing thickness as shown in the case of Pb and In. However, in the recently emerging 2D superconductors, some materials, *e.g.*, $MoS_2$, ZrNCl and $Bi_2Sr_2CaCu_2O_{8+x}$, exhibit almost the same $T_c$ as compared with that of bulk forms, and surprisingly, the $La_{2-x}Sr_xCuO_4/La_2CuO_{4+\delta}$ interfaces and FeSe thin films show much higher $T_c$ than that of bulk. Also, $KTaO_3$ is an example of superconductors which has not been known to superconduct in bulk and was discovered for the first time by the EDLT. Each mechanism is still in debate, but it seems that the origins are strongly material-specific. As demonstrated by FeSe, the highly-crystallized 2D materials might be promising candidates for realizing high-$T_c$ superconductivity.

For a future perspective, fabricating new superconducting systems including 2D materials and *in-situ* characterization combining various nanotechnologies should be one of the major streams. This is obvious when we see that the combined system of MBE-ARPES-STM are very popular all over the world. However, we would like to stress that the technology which will be utilized does not have to be expensive high technologies, as exemplified by the mechanical exfoliations and EDLT. Further attempts to introduce new technologies for fabrication and characterization of 2D or nanostructured superconductors with high crystallinities may be highly demanded.

Another important trend to be noted is that such 2D or nanostructured superconductors fit the first-principle calculations pretty well, in sharp contrast to the conventional 2D superconductors with disordered structures where realistic and precise band structures are not available, and thus the phenomenological or very general theories were dominating. However, in highly-crystalline structures, researchers are able to conduct the realistic calculations which will be powerful for pushing forward the 2D superconductors researches. In this sense, more material specific and precise models may be required. These signatures are already seen in the new 2D superconductors as introduced in this article.

**Acknowledgements**

We thank J. T. Ye, Y. Kasahara, Y. Kohama, M. Tokunaga, Y. Nakagawa and M. Onga for experimental cooperation, and Y. Nakamura, Y. Yanase and M. S. Bahramy for theoretical discussions. Y.S. was supported by the Japan Society for the Promotion of Science (JSPS) through a research fellowship for young scientists. This work was supported by the Strategic International Collaborative Research Program (SICORP-LEMSUPER) of the Japan Science and Technology Agency, Grant-in-Aid for Specially Promoted Research (no. 25000003) from JSPS and Grant-in Aid for Scientific Research on Innovative Areas (no. 22103004) from MEXT of Japan.

**Competing interests statement**

The authors declare no competing interests.

**Figure legends**

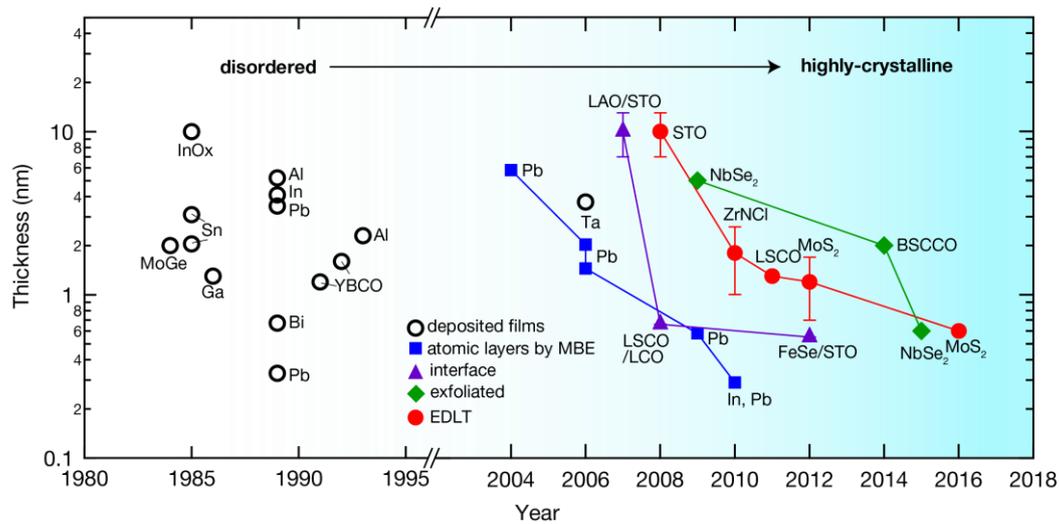

**Figure 1.   History of 2D superconductors.**
A plot of their thickness *vs.* year for 2D superconductors after 1980. In the last century, the majority of 2D superconductors were mostly fabricated by deposition of metallic thin films, which are strongly disordered, amorphous or granular (open circles)[6–13]. In the 21$^{st}$ century, atomic layers grown by MBE (blue squares)[16–20], interfacial superconductors (purple triangles)[21–23], exfoliated atomic layers (green diamonds)[24–27] and EDLT (red circles)[28–31,131] have been fabricated. All of them are highly-crystalline, in marked contrast with the conventional ones. Open circles includes three kinds of deposited thin films: InO$_x$, MoGe and Ta are sputtered thin films, Sn, Ga, Al, In, Pb and Bi are MBE-grown thin films, and YBa$_2$Cu$_3$O$_y$ (YBCO) is deposited by reactive evaporation.

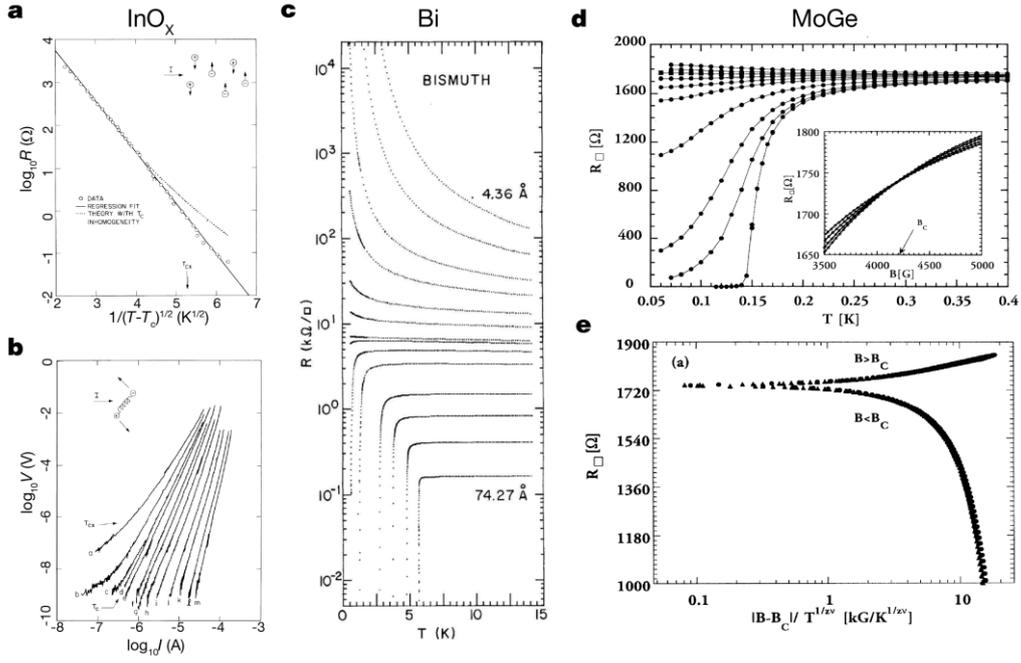

**Figure 2. 2D superconductivity in deposited metallic thin films.**
**a,** Plot of the logarithm of the resistance versus $(1-T/T_c)^{-1/2}$ in amorphous $InO_x$[45]. Here, $T_c$ is treated as the BKT transition temperature. These data represent the vortex-flow resistance of thermally excited vortices and antivortices. Inset: A schematic image of vortex flow process. **b,** Voltage-current characteristics in amorphous $InO_x$ thin films between 1.460 K and 1.939 K. Inset: A schematic image of nonlinear pair-breaking process due to unbinding of a vortex-antivortex pair. **c,** Evolution of superconductor-insulator transition in amorphous Bi films with the decrease in film thickness corresponding to the increase in degree of disorder[10]. **d,** Sheet resistance as a function of temperature in various magnetic fields in amorphous MoGe. Inset: Magnetoresistance around the critical magnetic field $B_c = 4.19$ kG[49]. **e,** Finte size scaling of sheet resistance plotted as a function of $|B-B_c|/T^{1/z\nu}$. Figures reproduced with permission from: APS.

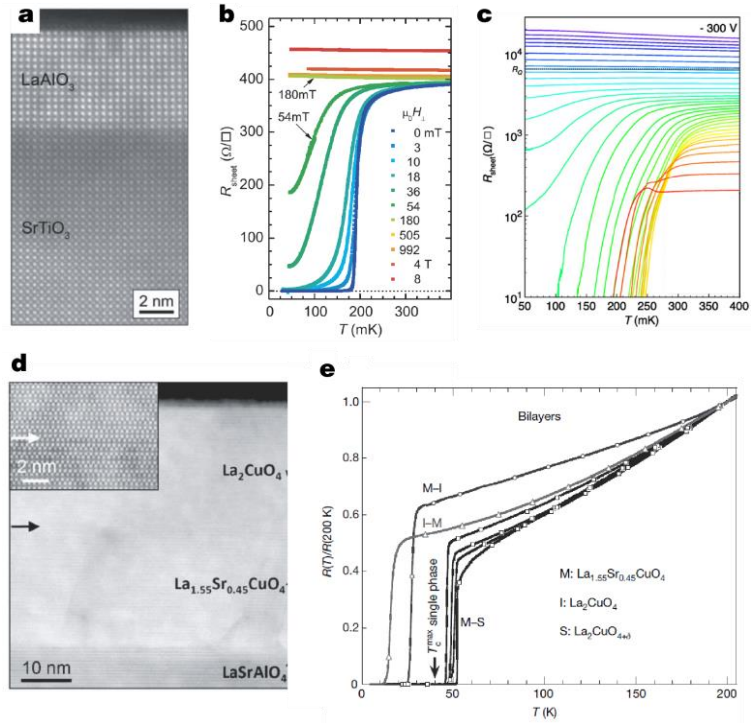

**Figure 3. Superconductivity in oxide interfaces.**
**a,** High-angle annular dark field scanning transmission electron microscopy image of a 15-unit-cell-thick $LaAlO_3$ film grown on $SrTiO_3$ [21]. **b,** Sheet resistance as a function of temperature for various magnetic fields. **c**, Superconductor-metal-insulator transition in $LaAlO_3/SrTiO_3$ controlled by electric fields[67]. **d**, Annular dark field image of the structure of $La_2CuO_4/La_{1.55}Sr_{0.45}CuO_4/LaSrAlO_4$ [22]. Inset: a magnified image of the M–I interface (white arrowed). **e**, $R(T)/R(200\ K)$ for various bilayers (M-I, I-M and M-S sequences). I is $La_2CuO_4$, vacuum-annealed and insulating; S is $La_2CuO_{4+\delta}$, oxygen-doped by annealing in ozone and superconducting; M is $La_{1.55}Sr_{0.45}CuO_4$, overdoped and metallic but not superconducting. Figures reproduced with permission from: AAAS and Nature Publishing Group.

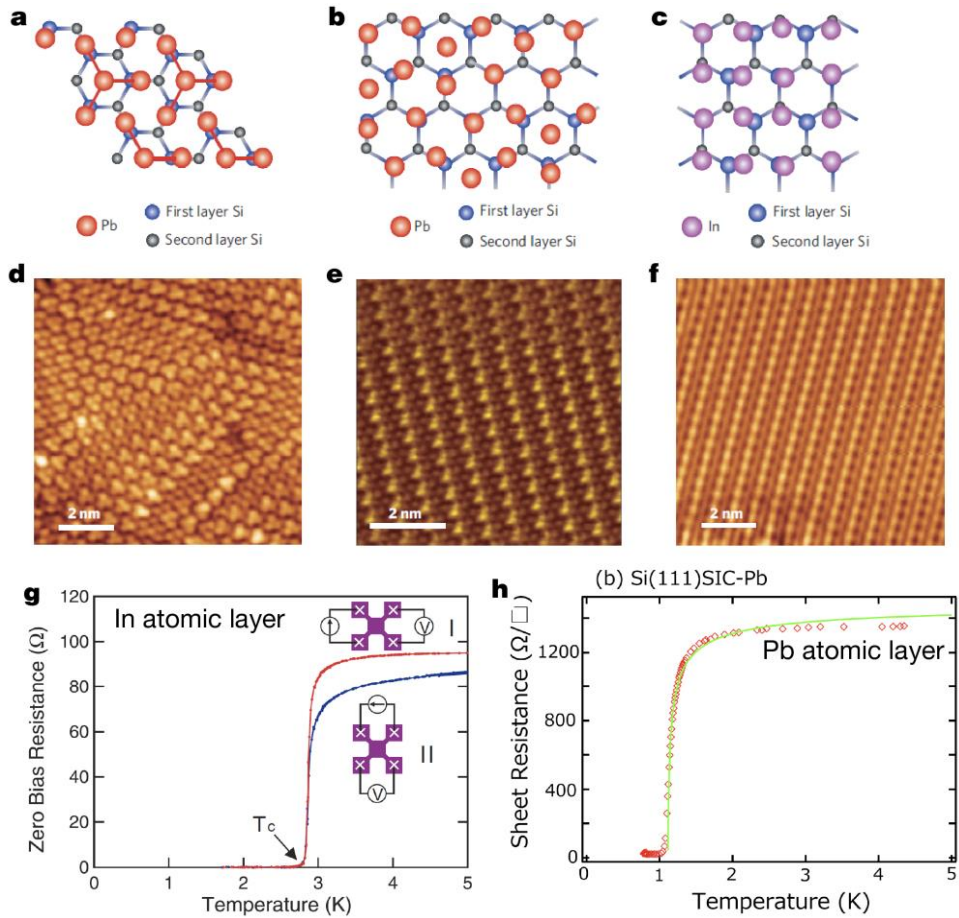

**Figure 4. Superconducting atomic layers of Pb and In grown by MBE.**
Schematic structure models (**a**‑**c**) and high-resolution STM images (**d**‑**f**) of the striped incommensurate -Pb (**a,d**), $\sqrt{7}\times\sqrt{3}$-Pb (**b,e**) and $\sqrt{7}\times\sqrt{3}$-In (**c,f**) phases grown on a Si(111) substrate [20]. Sheet resistance as a function of temperature in In (**g**) and Pb (**h**) atomic layers [83,84]. Insets in Fig. 4g: schematic drawings of the probe configurations I and II. Figures reproduced with permission from: Nature Publishing Group.

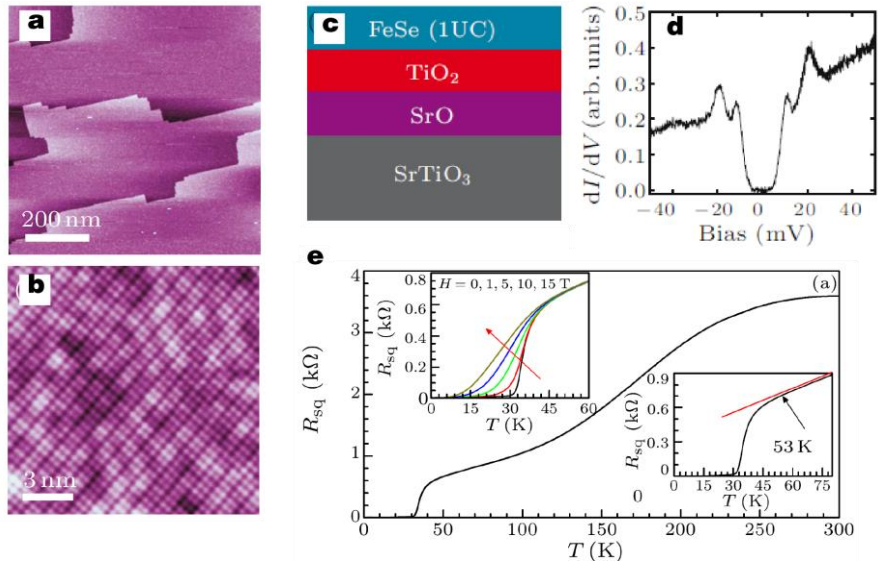

**Figure 5. High-temperature superconductivity in FeSe atomic layers.**
**a**, STM topography of SrTiO$_3$ substrate surface and **b**, atomically resolved Se terminated FeSe atomic layer [23]. **c**, A schematic of side-view of a FeSe film on the SrTiO$_3$ substrate. **d**, Tunneling spectrum taken on the 1-unit-cell-thick FeSe film on SrTiO$_3$ at 4.2 K. **e**, Temperature dependence of the sheet resistance of a 5-unit cell-thick FeSe film on insulating SrTiO$_3$ surface from 0 to 300 K. Here, the authors mentioned that the bottom first unit cell contribute to the superconductivity. Upper inset: Sheet resistance as a function of temperature in various out-of-plane magnetic fields. Lower inset: the magnification view between 0 to 80 K. Figures reproduced with permission from: IOP Science.

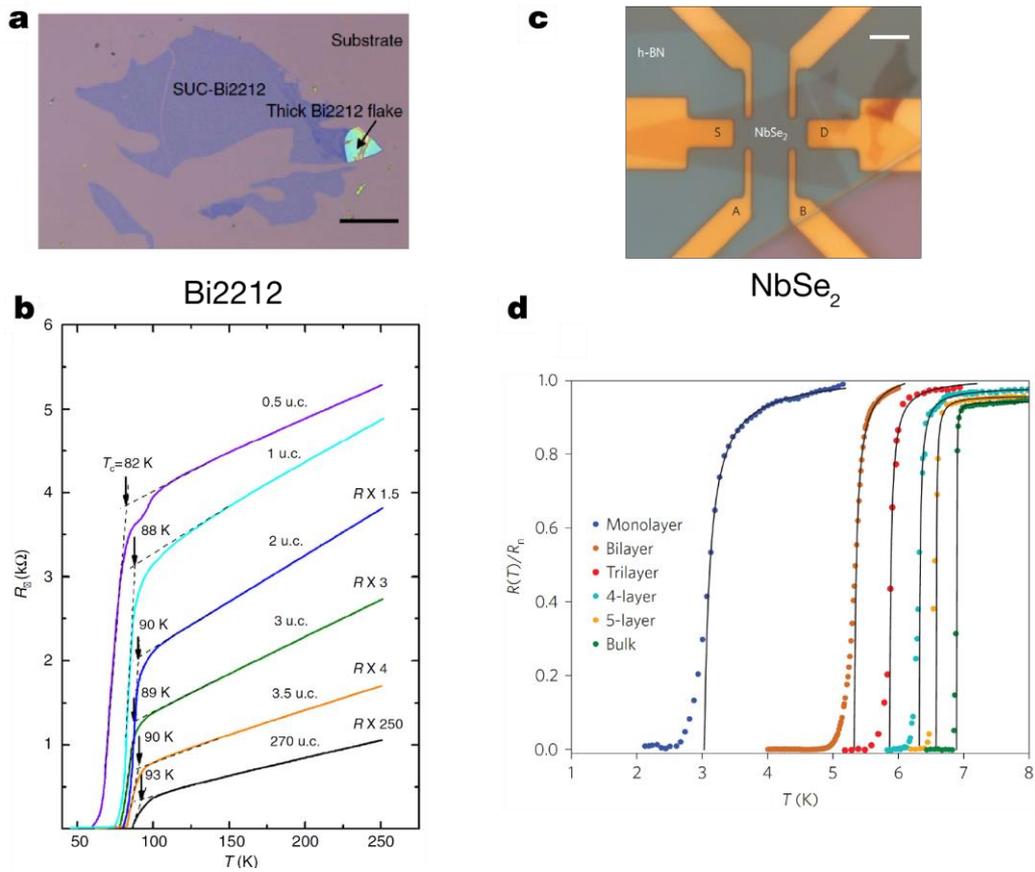

**Figure 6. Atomically thin superconductors based on exfoliated 2D crystals.**
**a**, Optical microscope image of several $Bi_2Sr_2CaCu_2O_{8+x}$ (Bi2212) flakes[25]. **b**, Temperature dependent sheet resistance for Bi2212 with various thicknesses from 270-unit-cell thick to half-unit-cell thick. **c**, Optical image of a bilayer $NbSe_2$ device capped by a thin h-BN layer for environmental protection.[27, 197]. **d**, Thickness variation of resistive superconducting transition of $NbSe_2$ atomic layers and bulk. Figures reproduced with permission from: Nature Publishing Group.

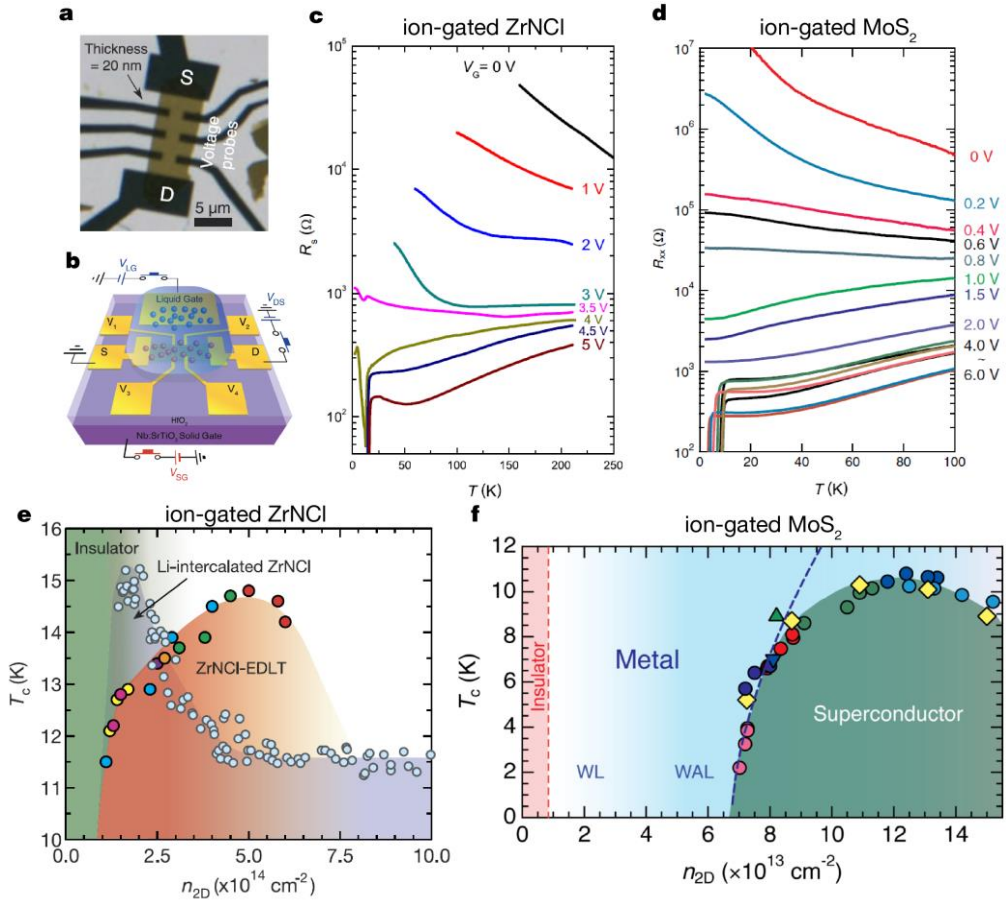

**Figure 7. Electric-field-induced superconductivity in 2D crystals.**
**a,** An optical image of a MoS$_2$ thin flake EDLT device before putting an ionic liquid. **b**, A schematic of an EDLT device. **c**, **d**, Insulator-to-superconductor transitions in ion-gated ZrNCl (**c**)[29] and MoS$_2$ (**d**)[30]. **e**, **f**, Electronic phase diagrams of ZrNCl (**e**)[137] and MoS$_2$ (**f**). Figures reproduced with permission from:AAAS.

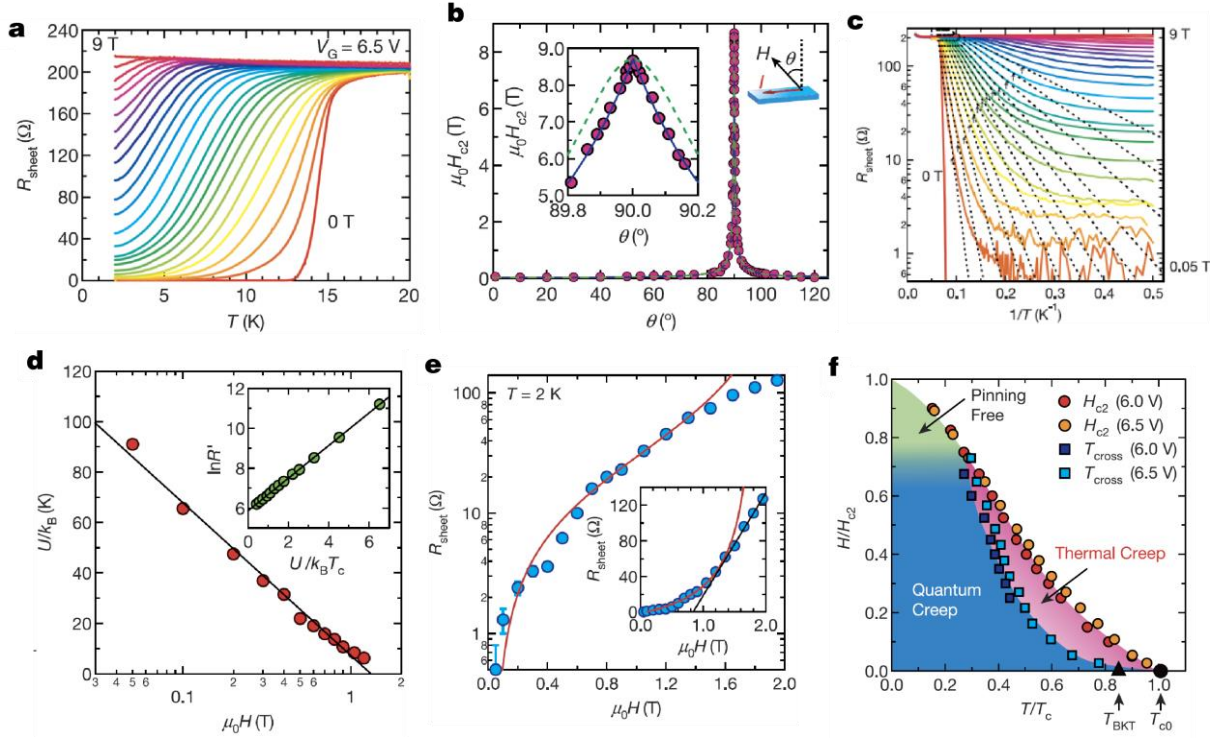

**Figure 8. Two-dimensionality and metallic ground state in ion-gated ZrNCl.**
**a**, Sheet resistance $R_{sheet}$ of a ZrNCl-EDLT as a function of temperature in various out-of-plane magnetic fields. **b**, Angular dependence of upper critical field. The inset shows a close-up of the region around $\theta = 90°$. **c**, Arrhenius plot of the sheet resistance of the ZrNCl-EDLT for different magnetic fields. The dashed lines are fitting for the data at high temperatures by $R_{sheet}(T\,H) = R'\exp(-U(H)/k_B T)$. **d**, Activation energy $U(H)$ is shown on a semi-logarithmic plot as a function of magnetic field. Inset: the same data plotted as $\ln R'$ versus $U(H)/k_B T_c$. **e**, Low-temperature saturated values of $R_{sheet}$ and fitting curve based on quantum creep model[166] as a function of magnetic field at 2 K. Inset: $H$-linear dependence of $R_{sheet}$ above 1.3 T (black solid line). **f**, Vortex phase diagram of the ZrNCl-EDLT[137]. Figures reproduced with permission from: AAAS

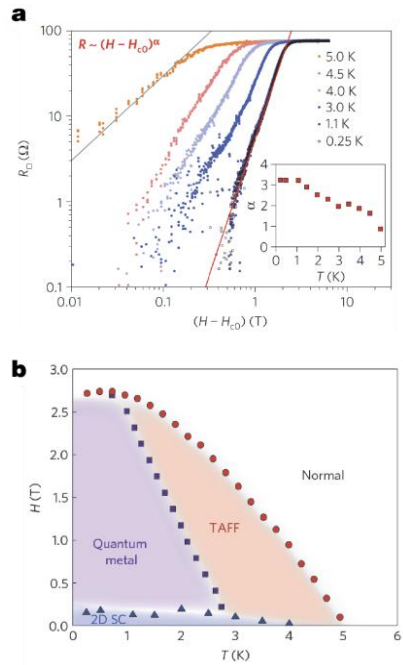

**Figure 9. Quantum metallic state in a bilayer NbSe$_2$ crystal.**
**a**, Logarithmic plot of magnetoresistance as a function of $(H - H_{c0})$ below the superconducting transition for different temperatures. Here, $H_{c0}$ is the characteristic field of true zero resistance state **b**, Magnetic field–temperature phase diagram of the bilayer NbSe$_2$ device showing the thermally assisted flux flow (TAFF) regime (red), Bose metal regime (purple) and 2D superconducting regime (blue)[144]. Figures reproduced with permission from: Nature Publishing Group.

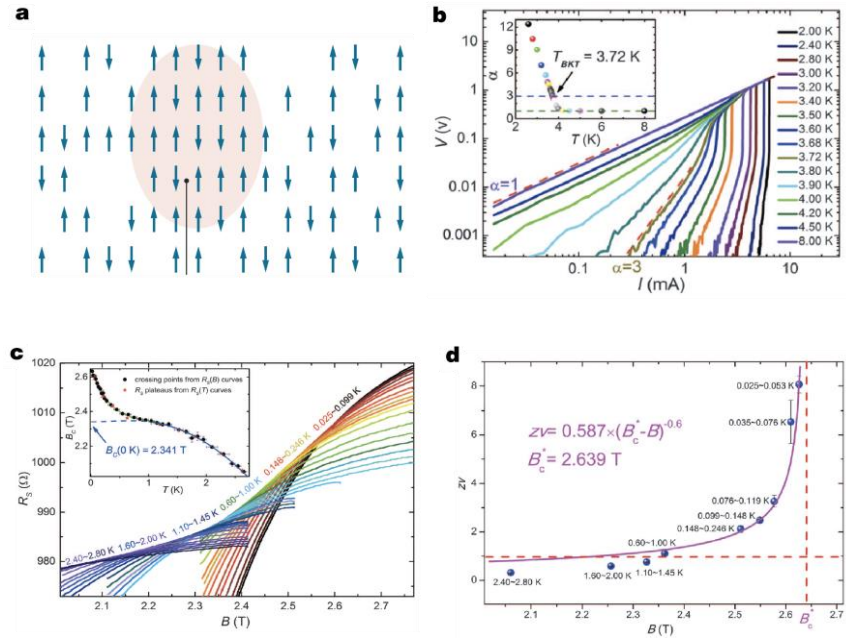

**Figure 10. Quantum Griffiths singularity in a superconducting Ga crystalline thin film.**
**a**, A schematic image of a rare ordered region in a model of 2D lattice of spins[204]. **b,** Voltage-current ($V$ - $I$) characteristic showing a BKT transition. Inset: the variation of exponent $\alpha$ as a function of temperature which is extracted from the power-law fittings for $V – I$ characteristic; $T_{BKT}$ = 3.72 K is defined by $\alpha$ =3 **c,** Magnetoresistance isotherm showing multiple crossing points. Inset: the critical magnetic fields $B_c$ as a function of temperature. Crossing points of $R_{sheet}(B)$ curves at every two adjacent temperatures are denoted as black dots on the transition boundary; the red stars come from the temperature plateaus on $R_{sheet}(T)$ curve. The blue line is the fitting curve Based on Werthamer-Helfand-Hohenberg theory. **d,** Critical exponents as a function of magnetic field showing quantum Griffiths singularity[175]. Figures reproduced with permission from AAAS.

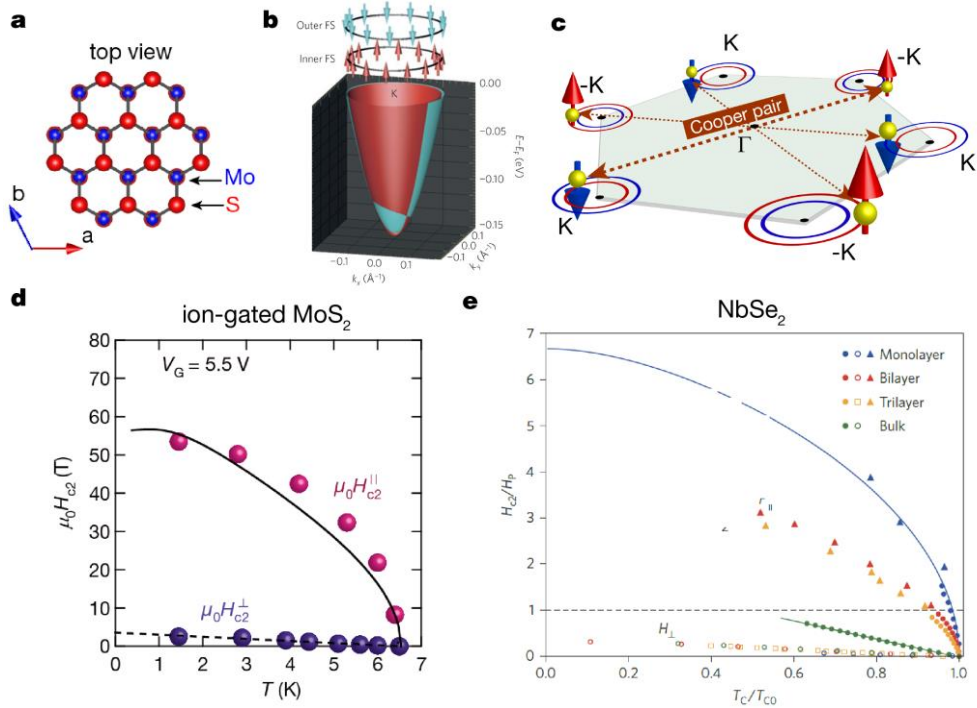

**Figure 11. Superconductivity protected by spin-valley locking in ion-gated MoS$_2$ and NbSe$_2$ bilayer.**
**a,** Top view of a monolayer MoS$_2$ crystal. **b,** Energy band dispersion and spin texture of the conduction band around the K point. Inner Fermi surface (FS) and outer FS at the K points have out-of-plane spin-polarization with up and down directions, respectively, due to effective valley Zeeman fields. **c,** Schematic image of the Fermi surfaces with valley-dependent spin polarization in the in-plane magnetic field geometry. The direction of each spin is orthogonal to the magnetic field. Inter-valley Cooper pairs between electrons in the K and -K valleys are protected by spin-valley locking effect, resulting in the superconductivity robust against an external magnetic field. **d, e,** In-plane upper critical field versus temperature in ion-gated MoS$_2$ (**d**)[195] and NbSe$_2$ (**e**)[197]. The maximum $H_{c2}^{\parallel}$ of 52 T at 1.5 K in ion-gated MoS$_2$ is over 4 times larger than usual Pauli limit of $1.86T_c$. The black solid curve and dashed line in Fig.11d show the numerically calculated in-plane upper critical field values assuming 10 % Rashba component of the total spin polarization, and out-of-plane upper critical fields based on 2D GL model. Figures reproduced with permission from Nature Publishing Group.
53